\documentclass[cits]{PoS}

\usepackage{aas_macros}

\title{NuGrid: $s$ process in massive stars}

\ShortTitle{NuGrid: $s$ process in massive stars}

\author{
\speaker{R. Hirschi}$^{abc}$,
U. Frischknecht$^{d}$,
M. Pignatari$^{abe}$,
F.-K. Thielemann$^{d}$,
M. E. Bennett$^{ab}$,
S. Diehl$^{afg}$,
C. L. Fryer$^{af}$,
F. Herwig$^{abh}$,
A. Hungerford$^{ag}$,
G. Magkotsios$^{aei}$,
G. Rockefeller$^{ag}$,
F. X. Timmes$^{ai}$,
and P. Young$^{ai}$\\
\llap{$^a$}The NuGrid Collaboration\\
\llap{$^b$}Astrophysics Group, Keele University, ST5 5BG, UK\\
\llap{$^c$}IPMU, University of Tokyo, Kashiwa, Chiba 277-8582, Japan\\
\llap{$^d$}Theoretical Astrophysics Group, University of Basel, Basel, 4056, Switzerland\\
\llap{$^e$}Joint Institute for Nuclear Astrophysics, University of Notre Dame, IN, 46556, USA\\
\llap{$^f$}Theoretical Astrophysics Group (T-6), Los Alamos National Laboratory, Los Alamos, NM, 87544, USA\\
\llap{$^g$}Computational Methods (CCS-2), Los Alamos National Laboratory, Los Alamos, NM, 87544, USA\\
\llap{$^h$}Dept. of Physics \& Astronomy, Victoria, BC, V8W 3P6, Canada\\
\llap{$^i$}School of Earth and Space Exploration, Arizona State University, Tempe, AZ 85287, USA\\
E-mail:\email{r.hirschi@epsam.keele.ac.uk}
}

\abstract{The $s$-process production in massive stars at very low
metallicities is expected to be negligible due to the low abundance of the neutron
source $^{22}$Ne, to primary neutron poisons
and decreasing iron seed abundances. However, recent models of massive stars
including the effects of rotation show that a strong production of 
$^{22}$Ne is possible in the helium core, as a consequence of the primary
nitrogen production (observed in halo metal poor stars). 
Using the PPN post-processing code, we studied the 
impact of this primary $^{22}$Ne on the $s$ process. We find 
a large production of $s$ elements between strontium and 
barium, starting with the amount of primary $^{22}$Ne predicted by stellar
models. There are several key reaction rate uncertainties influencing the 
$s$-process efficiency. 
Among them, $^{17}$O$(\alpha,\gamma)$ 
may play a crucial role strongly influencing the $s$ process efficiency, 
or it may play a negligible role, according to the rate used in the
calculations. 
We also report on the development of a new parallel (MPI) post-processing
code (MPPNP) designed to follow the complete nucleosynthesis in stars on highly 
resolved grids. We present here the first post-processing run from the ZAMS up
to the end of helium burning for a 15 $M_\odot$ model.} 

\FullConference{10th Symposium on Nuclei in the Cosmos\\
		 July 27 - August 1 2008\\
		 Mackinac Island, Michigan, USA}

\begin{document}

\section{Introduction}
Massive stars are known to produce elements heavier than the iron group
via rapid neutron captures during their explosion, $r$ process (see for example
the contribution by Qian and Kratz et al. 2007 \cite{KFP07}) and also via slow neutron captures ($s$ process) during the pre-supernova
evolution, forming the so-called weak $s$ component.
The weak $s$ process in massive stars with initial solar like composition
is well understood. $^{22}$Ne is the main neutron source and it is
produced starting from the initial CNO isotopes (The et al. 2000 and 2007
\cite{2000ApJ...533..998T,2007ApJ...655.1058T}, Raiteri et al. 1991
\cite{1991ApJ...371..665R,1991ApJ...367..228R}, Pignatari et al. in prep.). 
The weak $s$ process, producing mostly elements in the atomic mass range
$60\lesssim A \lesssim 90$, starts
at the end of helium burning when the temperature is high enough
to activate $^{22}$Ne$(\alpha,n)^{25}$Mg. More massive stars reach higher
temperatures at the end of He-burning and therefore burn more $^{22}$Ne. 
Consequently $s$-process
production during central helium burning increases with increasing stellar
mass. 
The $^{22}$Ne left over from helium burning is the main neutron source during
the subsequent carbon shell burning.
The carbon shell $s$-process
contribution depends on the history of
convective zones after the He-core burning and on different nuclear uncertainties 
(e.~g. $^{12}$C$(\alpha,\gamma)^{16}$O). 
The standard $s$-process production in massive stars depends on the initial 
metallicity. At low metallicity, the low iron seed abundance, the low
$^{22}$Ne content and the increasing strength of primary neutron poisons
limits the $s$-process efficiency, permitting only negligible
production of $s$ elements (e.~g. Raiteri et al. 1992
\cite{1992ApJ...387..263R}). 

\section{Weak $s$ process at low metallicity in rotating stars}
At solar metallicity, 
the main effect of
rotation on the $s$-process production is the enlargement of convective helium core due to additional
mixing and therefore a behaviour like non-rotating more massive stars
\cite{2005A&A...433.1013H}. Thus a 25 M$_\odot$ star with rotation behaves
 like non-rotating stars with masses between 30 and 40 M$_\odot$. 
Hence the $s$-process efficiency in He-core burning is enhanced
in rotating stars (Frischknecht et al. in prep.). 
 
At low metallicity, the impact of rotation is more important. 
Indeed, at the start of core He-burning,
carbon and oxygen are mixed upward into hydrogen rich regions leading to
a strong production of nitrogen 
(see Meynet et al. 2006 \cite{MEM06} and Hirschi 2007 \cite{2007A&A...461..571H}). 
Part of this primary nitrogen may enter
the convective He core and be transformed into primary $^{22}$Ne
by $\alpha$-captures. As a consequence, with respect to the
non-rotating models, the $^{22}$Ne available in the He-core
is strongly enhanced. According to Hirschi 
(2008 \cite{2008IAUS..250..217H}), about 1\% in mass of the helium
core is composed of $^{22}$Ne at
 the $s$-process activation.
\begin{figure}[h]
 \includegraphics[width=\textwidth]{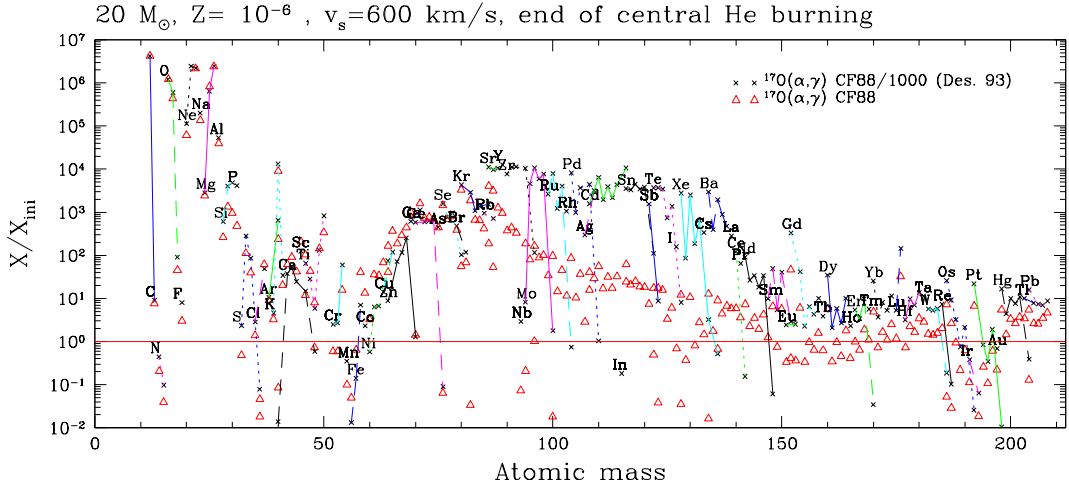}
 \caption{The overproduction factors after He-burning in the one-zone
 post-processing calculations.  
 Using a low $^{17}$O$(\alpha,\gamma)$ rate (black crosses) leads to a 
 strong increase of $s$-process overproduction between Sr and Ba.
 Isotopes with $X_{i}/X_{\rm ini}$ below the lower limit are not plotted.}
 \label{fig:20MolowZ_heburn}
\end{figure}

We present in Fig. \ref{fig:20MolowZ_heburn} one-zone 
post-processing runs up to the end of He-burning calculated with the PPN
code (see next Sect.) with an initial
metallicity of Z$=10^{-6}$. 
In order to reproduce the effect of rotational mixing on the helium
burning core composition in the one-zone calculation, 
we replaced 1\% in mass of $^4$He by $^{22}$Ne at the start of helium burning.
The primary $^{22}$Ne enhances the $s$ process compared to the non-rotating
case, where negligible amounts of $s$ elements are produced. 
The highest nucleosynthesis 
efficiency is around Sr with overproduction
factors ($X_{i}/X_{\rm ini}$) between thousand and ten thousand. As can be seen in Fig.
\ref{fig:20MolowZ_heburn}, 
iron seeds and in general elements lighter than strontium feed the $s$ nucleosynthesis
in the mass region between strontium (Sr) and barium (Ba). Beyond Ba, 
the $s$ efficiency rapidly falls, depending on the total neutron exposure.
The major neutron poisons are $^{16}$O, $^{25}$Mg and $^{22}$Ne, where 
$^{16}$O is the strongest neutron absorber. Whether or not $^{16}$O is an efficient 
poison depends on the ratio of $^{17}$O$(\alpha,\gamma)$ to $^{17}$O$(\alpha,n)$. 
According to the study of Descouvemont (1993 \cite{De93}), the $(\alpha,\gamma)$
channel should be orders of magnitude weaker than the $(\alpha,n)$ channel, in which case 
the neutrons captured by $^{16}$O are recycled by $^{17}$O$(\alpha,n)$. 
On the other hand, using the rates of Caughlan and Fowler (1988 \cite{CF88}),
$^{17}$O$(\alpha,\gamma)$ is about a factor ten slower than $^{17}$O$(\alpha,n)$
and a significant fraction of neutrons
captured by $^{16}$O are not re-emitted. In this case, $^{16}$O is the strongest 
neutron poison.
In Fig. \ref{fig:20MolowZ_heburn}, we show the importance of the
$^{17}$O$(\alpha,\gamma)$ rate by comparing the isotopic distributions
obtained using the rate of Caughlan and Fowler (1988
\cite{CF88}) (red triangles) and using this same rate
divided by a factor 1000 to reproduce the
$(\alpha,\gamma)/(\alpha,n)$ ratio suggested by Descouvemont (1993
\cite{De93}) (black crosses). The different $s$-process production between the two
calculations demonstrates the importance of the
$^{17}$O$(\alpha,\gamma)$ to $^{17}$O$(\alpha,n)$ ratio
for the $s$ process at low metallicity. 
This was also suggested by Rayet and Hashimoto (2000 \cite{RH00}) in standard $s$-process
calculations in massive stars at low metallicity. However, because of the large 
primary $^{22}$Ne production in rotating stars, in the present calculations 
the impact
of the $^{17}$O$(\alpha,\gamma)$ to $^{17}$O$(\alpha,n)$ ratio on the $s$-process efficiency 
is much stronger than in Rayet and Hashimoto (2000 \cite{RH00}).
A better knowledge of these two rates 
at He-burning temperature is highly desirable in order to obtain more reliable
$s$-process calculations at very low metallicity.
The strong production of $s$ elements between Sr and Ba is in agreement with 
Pignatari et al. (2008 \cite{2008ApJ..Pignatari..sproc}), where the 
$^{17}$O$(\alpha,\gamma)$ rate of Descouvemont 1993 \cite{De93} is used
 and where the amount of primary $^{22}$Ne is in agreement with 
Hirschi 2008 \cite{2008IAUS..250..217H}.
The boosted $s$ process
due to primary $^{14}$N production may provide a new 
$s$-process component with important implications for nucleosynthesis
at low metallicity. Massive rotating stars may therefore contribute considerable
amounts of isotopic abundances between Sr and Ba to the Galactic
chemical evolution at halo metallicities, which could provide a
possibility to explain the high Sr enrichment and the high Sr/Ba ratio
(see Pignatari et al. 2008 \cite{2008ApJ..Pignatari..sproc} for more details). 
In order to make a quantitative and more precise statement about the
importance of this $s$ process occurring in rotating low-metallicity
stars, further investigations are needed. 

\section{Multi-zone parallel (MPI) post-processing code, MPPNP}
\begin{figure}[h]
 \includegraphics[width=\textwidth]{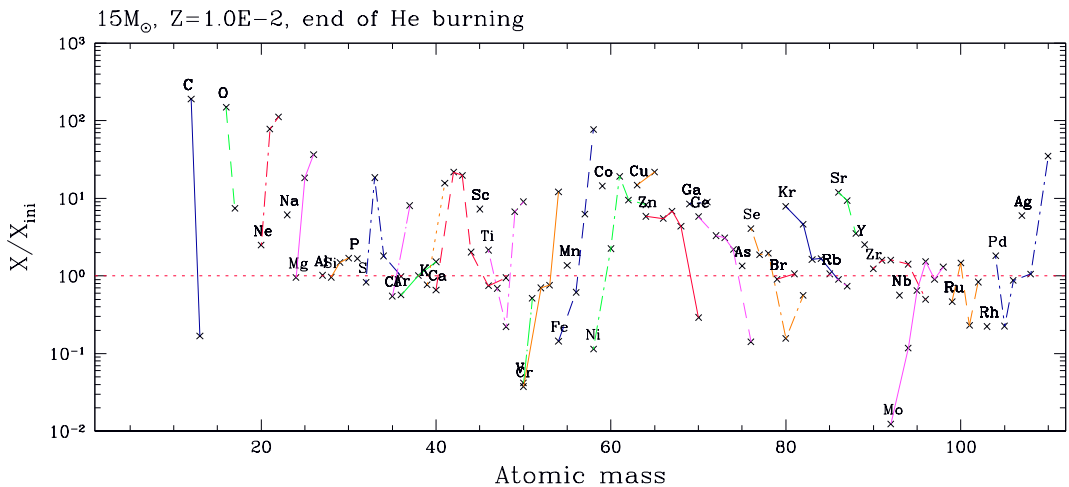}
 \caption{Overproduction factors in the convective core at the end of 
 He-burning. Isotopes with $X_{i}/X_{\rm ini}$ below the lower limit are not plotted.} 
 \label{fig:15Mo_heburn}
\end{figure}

Although only a few isotopes are crucial for the energy generation in
massive stars, many more are important for the nucleosynthesis, for example 
to determine how much $s$ process is made in massive stars.
Since it is not necessary to follow many of these species within a
stellar evolution calculation, we developed a
post-processing network, called PPN, that allows us to follow the
complete nucleosynthesis taking place in massive stars. It also enables
testing of the importance of various reaction rates and especially the use
of the same set of nuclear reactions in different stellar environments. 
The MPPNP
variant uses MPI and is therefore much faster than a serial code. Using
MPPNP, we have post-processed a full stellar evolution model of 15 M$_\odot$
at Z$=0.01$ calculated with the Geneva code \cite{GVAcode} from
the ZAMS up to the end of helium burning with a 400-isotope network up
to Ag. The overabundance pattern at the end of the core He-burning phase
is shown in Fig. \ref{fig:15Mo_heburn}. As expected, the weak $s$-process 
production in
a 15 $M_\odot$ star is modest, with overproduction factors up to 10 
for $s$-only isotopes between iron and strontium.
This is due to the low central temperature reached at the end of the core 
He-burning phase in a 15 $M_\odot$ star (compared to more massive stars) 
with a marginal activation of the $^{22}$Ne$(\alpha,n)^{25}$Mg during He-burning.
We are currently testing MPPNP in the advanced stages and we plan to calculate
the full nucleosynthesis for a large range of masses and metallicities.
The MPPNP code will also be able to post-process AGB models 
(see contribution by Pignatari) and another variant of PPN, called TPPNP 
will follow  trajectories of multi-dimensional simulations of 
supernova explosion and convective-reactive events in stars  
(see contribution by Herwig).

\providecommand{\href}[2]{#2}\begingroup\raggedright\endgroup

\end{document}